\documentclass{article}

\usepackage{microtype}
\usepackage{graphicx}
\usepackage{booktabs} %
\usepackage[hyphens]{url}
\usepackage{ragged2e}
\usepackage{adjustbox}
\usepackage[nointegrals]{wasysym}
\usepackage{subcaption}
\newcommand{\wcircle}{{$\RIGHTcircle$}}
\newcommand{\bcircle}{{$\CIRCLE$}}
\newcommand{\ecircle}{{$\Circle$}}

\newcommand*\rot{\rotatebox{25}}

\usepackage{hyperref}

\usepackage[accepted]{icml2024-arxiv}

\usepackage{amsmath}
\usepackage{amssymb}
\usepackage{mathtools}
\usepackage{amsthm}

\usepackage[capitalize,noabbrev]{cleveref}

\theoremstyle{plain}
\newtheorem{theorem}{Theorem}[section]
\newtheorem{proposition}[theorem]{Proposition}
\newtheorem{lemma}[theorem]{Lemma}
\newtheorem{corollary}[theorem]{Corollary}
\theoremstyle{definition}
\newtheorem{definition}[theorem]{Definition}
\newtheorem{assumption}[theorem]{Assumption}
\theoremstyle{remark}

\usepackage{booktabs} %
\usepackage{tabularx}
\usepackage{adjustbox}
\usepackage{framed}
\usepackage{enumitem}
\usepackage{ragged2e}
\newcolumntype{P}[1]{>{\RaggedRight\hspace{0pt}}m{#1}}
\usepackage{color, colortbl}
\definecolor{Gray}{gray}{0.94}
\usepackage{pifont}
\usepackage[capitalize]{cleveref}
\usepackage{tcolorbox}
\usepackage{xspace}
\usepackage{xcolor}
\tcbuselibrary{skins}
\usepackage{caption}
\usepackage{mwe}
\usepackage{caption}
\usepackage{subcaption} %
\usepackage{tikz}
\usepackage{float}

\def\Snospace~{\S{}}

\usepackage[textsize=tiny]{todonotes}

\icmlauthorrunning{Kapoor and Bommasani et al.}

\newcommand\tldr[1]{\textcolor{gray}{\textit{[#1]} \\}}
\newcommand\tldrDone[1]{}
\newcommand\todocite[1]{\textcolor{green}{[Add cite: #1]}}
\newcommand\ska[1]{\textcolor{orange}{[SK: #1]}}
\newcommand\rb[1]{\textcolor{magenta}{[RB: #1]}}
\newcommand\kk[1]{\textcolor{olive}{[KK: #1]}}
\newcommand\srl[1]{\textcolor{violet}{[SL: #1]}}
\newcommand\an[1]{\textcolor{blue}{[AN: #1]}}
\newcommand\pl[1]{\textcolor{purple}{[PL: #1]}}

\renewcommand\tldr[1]{}
\renewcommand\tldrDone[1]{}
\renewcommand\todocite[1]{}
\renewcommand\ska[1]{}
\renewcommand\rb[1]{}
\renewcommand\kk[1]{}
\renewcommand\srl[1]{}
\renewcommand\an[1]{}
\renewcommand\pl[1]{}

\newcommand\eg{e.g.\xspace}

\ifthenelse{\isundefined{\definition}}{\newtheorem{definition}{Definition}}{}
\ifthenelse{\isundefined{\assumption}}{\newtheorem{assumption}{Assumption}}{}
\ifthenelse{\isundefined{\hypothesis}}{\newtheorem{hypothesis}{Hypothesis}}{}
\ifthenelse{\isundefined{\proposition}}{\newtheorem{proposition}{Proposition}}{}
\ifthenelse{\isundefined{\theorem}}{\newtheorem{theorem}{Theorem}}{}
\ifthenelse{\isundefined{\lemma}}{\newtheorem{lemma}{Lemma}}{}
\ifthenelse{\isundefined{\corollary}}{\newtheorem{corollary}{Corollary}}{}
\ifthenelse{\isundefined{\alg}}{\newtheorem{alg}{Algorithm}}{}
\ifthenelse{\isundefined{\example}}{\newtheorem{example}{Example}}{}

\icmltitlerunning{\thepage $\quad\cdot\quad$On the Societal Impact of Open Foundation Models}

\begin{document}

\twocolumn[
\icmltitle{
On the Societal Impact of Open Foundation Models}

\icmlsetsymbol{equal}{*}

\begin{icmlauthorlist}
\icmlauthor{Sayash Kapoor}{equal,princeton}
\icmlauthor{Rishi Bommasani}{equal,stanford}
\\\vspace{0.1cm}
\icmlauthor{Kevin Klyman}{stanford}
\icmlauthor{Shayne Longpre}{mit}
\icmlauthor{Ashwin Ramaswami}{georgetown}
\icmlauthor{Peter Cihon}{github}
\icmlauthor{Aspen Hopkins}{mit}
\\\vspace{0.1cm}
\icmlauthor{Kevin Bankston}{cdt,georgetown}
\icmlauthor{Stella Biderman}{eai}
\icmlauthor{Miranda Bogen}{cdt,princeton}
\icmlauthor{Rumman Chowdhury}{hi}
\icmlauthor{Alex Engler}{brookings}
\icmlauthor{Peter Henderson}{princeton}
\icmlauthor{Yacine Jernite}{hf}
\icmlauthor{Seth Lazar}{anu}
\icmlauthor{Stefano Maffulli}{osi}
\icmlauthor{Alondra Nelson}{ias}
\\
\icmlauthor{Joelle Pineau}{meta}
\icmlauthor{Aviya Skowron}{eai}
\icmlauthor{Dawn Song}{ucb}
\icmlauthor{Victor Storchan}{mozilla}
\icmlauthor{Daniel Zhang}{stanford}
\\\vspace{0.1cm}
\icmlauthor{Daniel E. Ho}{stanford}
\icmlauthor{Percy Liang}{stanford}
\icmlauthor{Arvind Narayanan}{princeton}
\\\vspace{0.3cm}
\icmlauthor{\textmd{February 27, 2024}}{}
\end{icmlauthorlist}

\icmlaffiliation{princeton}{Princeton University}
\icmlaffiliation{stanford}{Stanford University}
\icmlaffiliation{mit}{Massachusetts Institute of Technology}
\icmlaffiliation{georgetown}{Georgetown University}
\icmlaffiliation{github}{GitHub}
\icmlaffiliation{mit}{Massachusetts Institute of Technology}
\icmlaffiliation{cdt}{Center for Democracy and Technology}
\icmlaffiliation{eai}{Eleuther AI}
\icmlaffiliation{hi}{Humane Intelligence}
\icmlaffiliation{brookings}{Work done while at Brookings Institution}
\icmlaffiliation{hf}{Hugging Face}
\icmlaffiliation{anu}{Australian National University}
\icmlaffiliation{osi}{Open Source Initiative}
\icmlaffiliation{ias}{Institute for Advanced Study}
\icmlaffiliation{meta}{Meta}
\icmlaffiliation{ucb}{University of California, Berkeley}
\icmlaffiliation{mozilla}{Mozilla AI}

\icmlcorrespondingauthor{Sayash Kapoor}{sayashk@princeton.edu}
\icmlcorrespondingauthor{Rishi Bommasani}{nlprishi@stanford.edu}

\icmlkeywords{Machine Learning, ICML}

\vskip 0.3in
]

\printAffiliationsAndNotice{\icmlEqualContribution} %

\enlargethispage{2\baselineskip}
\begin{abstract}
Foundation models are powerful technologies: how they are released publicly directly shapes their societal impact.
In this position paper, we focus on \textit{open} foundation models, defined here as those with broadly available model weights (\eg Llama 2, Stable Diffusion XL).
We identify five distinctive properties (\eg greater customizability, poor monitoring) of open foundation models that lead to both their benefits and risks.
Open foundation models present significant benefits, with some caveats, that span innovation, competition, the distribution of decision-making power, and transparency.
To understand their risks of misuse, we design a risk assessment framework for analyzing their \textit{marginal risk}. 
Across several misuse vectors (\eg cyberattacks, bioweapons), we find that current research is insufficient to effectively characterize the marginal risk of open foundation models relative to pre-existing technologies. 
The framework helps explain why the marginal risk is low in some cases, clarifies disagreements about misuse risks by revealing that past work has focused on different subsets of the framework with different assumptions, and articulates a way forward for more constructive debate.
Overall, our work helps support a more grounded assessment of the societal impact of open foundation models by outlining what research is needed to empirically validate their theoretical benefits and risks.
\end{abstract}

\hypertarget{introduction}{\section{Introduction}}
\label{sec:introduction}

Foundation models \citep{bommasani2021opportunities} are the centerpiece of the modern AI ecosystem, catalyzing a frenetic pace of technological development, deployment, and adoption that brings with it controversy, scrutiny, and public attention. 
\textit{Open foundation models}\footnote{We define \textit{open foundation models} as foundation models with widely available model weights \citep[see][]{EO14110, ntia2024rfc}.} like BERT, CLIP, Whisper, BLOOM, Pythia, Llama 2, Falcon, Stable Diffusion, Mistral, OLMo, Aya, and Gemma play an important role in this ecosystem.
These models allow greater customization and deeper inspection of how they operate, giving developers greater choice in selecting foundation models.
However, they may also increase risk, especially given broader adoption, which has prompted pushback, especially around risks relating to biosecurity, cybersecurity, and disinformation.
How to release foundation models is a central debate today, often described as open vs. closed.

Simultaneously, policymakers are confronting how to govern open foundation models.
In the United States, the recent Executive Order on Safe, Secure, and Trustworthy Development and Use of Artificial Intelligence mandates that the Department of Commerce prepare a report for the President on the benefits and risks of open foundation models \citep{EO14110}.
In the European Union, open foundation models are partially exempt from obligations under the recently-negotiated AI Act.
And consideration of widely available model weights is a stated priority of the UK's AI Safety Institute \citep{ukaisi2023introducing}.

Given disagreement within the AI community and uncertainty on appropriate AI policy~(\autoref{sec:background}), our paper clarifies the benefits and risks of open foundation models.
We decompose the analysis of the societal impact of open foundation models into two steps.
First, we articulate five distinctive properties of open foundation models (\autoref{sec:distinctive_properties}).
Open foundation models are marked by broader access, greater customizability, the potential for local inference, an inability to rescind model access once released, and weaker monitoring.

Second, we outline how these distinctive properties lead to specific benefits and risks of open foundation models.
The benefits we identify are distributing decision-making power, reducing market concentration, increasing innovation, accelerating science, and enabling transparency~(\autoref{sec:benefits}).
We highlight considerations that may temper these benefits in practice (\eg model weights are sufficient for some forms of science, but access to training data is necessary for others and is not guaranteed by release of weights).

Turning to risks, we present a framework for conceptualizing the \textit{marginal risk} of open foundation models: that is, the extent to which these models increase societal risk by intentional misuse beyond 
closed foundation models or pre-existing technologies, such as web search on the internet~(\autoref{sec:risks}).
Surveying seven common misuse vectors described for open foundation models (\eg disinformation, biosecurity, cybersecurity, non-consensual intimate imagery, scams), we find that past studies do not clearly assess the marginal risk in most cases.

Our framework helps explain why the marginal risk is low in some cases where we already have evidence from past waves of digital technology (such as the use of foundation models for automated vulnerability detection in cybersecurity).
It also helps retrospectively explain why the research on the dangers of open foundation models has been so contentious---past studies implicitly analyze risks for different subsets of our framework. 
The framework provides a way to have a more productive debate going forward, by outlining the necessary components of a complete analysis of the misuse risk of open foundation models.
Namely, while the current evidence for marginal risk is weak for several misuse vectors, we encourage more empirically grounded work to assess the marginal risk, recognizing the nature of this risk will evolve as model capabilities and societal defenses evolve.

By clearly articulating the benefits and risks of open foundation models, including where current empirical evidence is lacking, we ground ongoing discourse and policymaking.
Specifically, we use our analysis to direct recommendations at AI developers, researchers investigating the risks of AI, competition regulators, and policymakers~(\autoref{sec:discussion}).
Action from these stakeholders can further clarify the societal impact of open foundation models and, thereby, enhance our ability to reap their benefits while mitigating risks.
\hypertarget{background}{\section{Background}}
\label{sec:background}

The release landscape for foundation models is complex \citep{sastry2021release, liang2022community-norms, solaiman2023gradient}.
In particular, several \textit{assets} exist (\eg the model, data, code): for each asset, there is the matter of \textit{who} can access the asset (\eg user restrictions like requiring that the user be 18 or older) and for \textit{what} purposes (\eg use restrictions that prohibit usage for competing against the model developer).\footnote{Models are often accompanied by licenses that specify these terms. The Open Source Initiative designates some licenses, generally applied to code, as open source and is in the process of leading a similar effort to define open source AI. See: \url{https://opensource.org/deepdive/}}
Further, the degree of access may \textit{change over time} (\eg staged release to broaden access, deprecation to reduce access).

In this paper, we consider a reductive, but useful, dichotomy between open and closed foundation models to facilitate analysis. 
We define \textit{open foundation models} as foundation models with widely available model weights.
(For simplicity, we refer to any non-open foundation model as \textit{closed}.)
In particular, with respect to the dimensions of release we describe, this means an open foundation model 
(i) must provide weights-level access, (ii) need not be accompanied by the open release of any other assets (\eg code, data, or compute), (iii) must be widely available, though some restrictions on users (\eg based on age) may apply, (iv) need not be released in stages, and (v) may have use restrictions. 
Our definition is consistent with the recent US Executive Order's notion of ``foundation models with widely available model weights" \citep{EO14110}.

We consider this dichotomy because many of the risks described for open foundation models arise because developers relinquish exclusive control over downstream model use once model weights are released. 
For example, if developers impose restrictions on downstream usage, such restrictions are both challenging to enforce and easy for malicious actors to ignore.
On the other hand, in the face of malicious use, developers of closed foundation models can, in theory, reduce, restrict, or block access to their models. 
In short, open release of model weights is irreversible. 

As a result, some argue that widely available model weights could enable better research on their effects, promote competition and innovation, and improve scientific research, reproducibility, and transparency \citep{toma_generative_2023, creative_commons_supporting_2023,cihon_how_2023, mozilla_joint_2023}.
Others argue that widely available model weights would enable malicious actors \citep{seger2023open,brundage_malicious_2018} to more effectively misuse these models to generate disinformation~\cite{solaiman2019release}, non-consensual intimate imagery~\cite{satter_fbi_2023, maiberg_inside_2023}, scams~\cite{hazell_large_2023}, and bioweapons~\cite{gopal_will_2023,soice2023large,sandbrink2023artificial,matthews_scientists_2023,service_could_2023,bray_ai_2023}. \autoref{sec:prior work} provides a brief history of the debate on open foundation models.
 
\section{Distinctive properties of open foundation models} \label{sec:distinctive_properties}
Our work aims to better conceptualize the benefits and risks of open foundation models, especially in light of widespread disagreement within and beyond the AI community.
Fundamentally, we decompose this into (i) identifying distinctive properties of open foundation models and (ii) reasoning about how those properties contribute to specific societal benefits and risks.
Here, we enumerate five distinctive properties of open foundation models compared to closed foundation models.
Note that other properties of foundation models, while not unique to open foundation models, may nonetheless influence the analysis of the benefits and risks of open foundation models.
In particular, as models become more capable \citep{anderljung_frontier_2023}, these capabilities are likely to present new beneficial market opportunities but also greater misuse potential (\eg more persuasive and targeted disinformation).

\textbf{Broader access.}
Given our definition, open foundation models require that the model weights be widely available, if not to the public as a whole.
While there may be some restrictions on who can use the model, given that such user restrictions are difficult to enforce or verify (as demonstrated by Meta's LLaMA 1 release in March 2023), model weights may effectively be available to the public. 
Functional barriers to use, ranging from requisite expertise to compute affordability, may nonetheless remain.

\textbf{Greater customizability.}
By releasing model weights, open foundation models are readily customized for various downstream applications.
Weights (and associated computational artifacts made available, such as activations and gradients) permit a wide range of adaptation methods for modifying the model, such as quantization~\cite{frantar_optq_2023}, fine tuning~\cite{zhang_llama-adapter_2023,dettmers2023qlora}, and pruning~\cite{xia_sheared_2023}.
While some closed foundation model developers permit certain adaptation methods (\eg OpenAI allows fine tuning of GPT 3.5 as of January 2024), these methods tend to be more restrictive, costly, and ultimately constrained by the model developer's implementation. 
The customizability of open foundation models prevents model alignment interventions from being effective---such as by allowing users to fine-tune away alignment interventions~\cite{narayanan_model_2023}, though similar issues also arise when closed models can be fine tuned~\cite{qi2023finetuning}.

\textbf{Local adaptation and inference ability.}
Users of an open foundation model can directly deploy it on local hardware,  which removes the need for transferring data to the model developer. 
This allows for the direct use of the models without the need to share sensitive data with third parties, which is particularly important in sectors where confidentiality and data protection are necessary---such as because of the sensitive nature of content or regulation around how data should be stored or transferred. 
This is important for applications of foundation models in domains such as healthcare and finance. 

\textbf{Inability to rescind model access.}
Once the weights for a foundation model are made widely available, little recourse exists for the foundation model developer to rescind access.
While the foundation model developer, in coordination with distribution channels used to share model weights, can stop further access, existing copies of the model weights cannot be revoked. 
Furthermore, despite the developer's objections, users can redistribute model weights through, for example, peer-to-peer distribution~\cite{vincent_metas_2023}. 

\textbf{Inability to monitor or moderate model usage.}
For open foundation models, inference may be performed (i) locally (\eg on a personal computer or self-owned cluster), (ii) on generic third-party computing platforms such as cloud services (\eg Google Cloud Platform, Microsoft Azure), or (iii) on dedicated model hosting platforms (\eg Together, Amazon Bedrock).
In all cases, foundation model developers do not observe inference by default, making monitoring or moderation challenging, especially for local inference. 
Since dedicated model hosts are aware of what models are being used, developers may be able to coordinate with hosts to implement certain forms of monitoring/moderation.
\hypertarget{benefits}{\section{Benefits of Open Foundation Models}}
\label{sec:benefits}

Having established distinctive properties of open foundation models, we now critically analyze key benefits for open foundation models that emerge from these properties.

\textbf{Distributing who defines acceptable model behavior.}
\textit{Broader access and greater customizability expand who is able to specify the boundary of acceptable model behavior.
}

Developers of closed foundation models exercise unilateral control in determining what is and is not acceptable model behavior.
Given that foundation models increasingly intermediate critical societal processes \citep[\eg access to information, interpersonal communication;][]{lazar2023algorithmiccity}, much as social media platforms do today, the definition of what is acceptable model behavior is a consequential decision that should take into account the views of stakeholders and the context where the model is applied.
In contrast, while developers may initially specify and control how the model responds to user queries, downstream developers who use open foundation models can modify them to specify alternative behavior.
Open foundation models allow for greater diversity in defining what model behavior is acceptable, whereas closed foundation models implicitly impose a monolithic view that is determined unilaterally by the foundation model developer. 

\textbf{Increasing innovation.}
\textit{Broader access, greater customizability, and local inference expand how foundation models are used to develop applications.
}

Since open foundation models can be more aggressively customized, they better support innovation across a range of applications.
In particular, since adaptation and inference can be performed locally, application developers can more easily adapt or fine-tune models on large proprietary datasets without data protection and privacy concerns. Similarly, the customizability of open models allows improvements such as furthering the state-of-the-art across different languages~\cite{pipatanakul_typhoon_2023}.
While some developers of closed foundation models provide mechanisms for users to opt out of data collection, the data storage, sharing, and usage practices of foundation model developers are not always transparent.

However, the benefits of open foundation models for innovation may have limits due to potential comparative disadvantages in improving open foundation models over time. For example, open foundation model developers generally do not have access to user feedback and interaction logs that closed model developers do for improving models over time.
Further, because open foundation models are generally more heavily customized, model usage becomes more fragmented and lessens the potential for strong economies of scale. However, new research directions such as merging models might allow open foundation model developers to reap some of these benefits (akin to open source software)~\cite{raffel_building_2023}.
More generally, the usability of foundation models strongly influences innovation \citep{vipra2023concentration}: factors beyond whether a model is released openly such as the capabilities of the model and the quality of potential inference APIs shape usability. 

\textbf{Accelerating science.}
\textit{Broader access and greater customizability facilitate scientific research. The availability of other key assets (especially training data) would further accelerate scientific research.}

Foundation models are critical to modern scientific research, within and beyond the field of artificial intelligence. 
Broader access to foundation models enables greater inclusion in scientific research, and model weights are essential for several forms of research across AI interpretability, security, and safety (see \autoref{tab:science}). 
Ensuring ongoing access to specific models is essential for the scientific reproducibility of research, something that has been undermined to date by the business practice of closed model developers to retire models regularly~\cite{kapoor_openais_2023}.
And since closed foundation models are often instrumented by safety measures by developers, these measures can complicate or render some research impossible.
For example, \citet{park_social_2022} use foundation models without safety filters because their research aims to simulate human behavior (including toxic speech). Most closed foundation models would suppress these outputs.

However, model weights alone are insufficient for several forms of scientific research.
Other assets, especially the data used to build the model, are necessary. 
For example, to understand how biases propagate, and are potentially amplified, requires comparisons of data biases to model biases, which in turn requires access to the training data \citep{wang_directional_2021}. 
Access to data and other assets, such as model checkpoints, has already enabled wide-ranging downstream research~\cite{tian2023joma,choi2023tools,longpre2023pretrainer}.
While some projects prioritize accessibility to such assets with the stated goal of advancing scientific research on foundation models~\cite{scao2022bloom,biderman2023pythia}, it is not common for open models in general.
In fact, even the basic validity of model's evaluation depends on some transparency about the training data. For example, issues such as contamination might lead to overoptimistic results on benchmarks~\cite{kapoor_promises_2024,narayanan_gpt-4_2023}. Access to information about the data can allow us to assess the amount of overlap between the training data and the test set.

\textbf{Enabling transparency.} 
\textit{Broad access to weights enables some forms of transparency. The availability of other key assets (such as documentation and training data) would further improve transparency.}

Transparency is a vital precondition for responsible innovation and public accountability.
Yet digital technologies are plagued by problematic opacity \citep[see][\S2.2]{bommasani2023foundation}.
Widely available model weights enable external researchers, auditors, and journalists to investigate and scrutinize foundation models more deeply.
In particular, such inclusion is especially valuable given that the foundation model developers often underrepresent marginalized communities that are likely to be subject to the harms of foundation models.
The history of digital technology demonstrates that broader scrutiny, including by those belonging to marginalized groups that experience harm most acutely, reveals concerns missed by developers \citep{sweeney2013discrimination, noble2018algorithms, buolamwini2018gender, raji2019actionable}. The 2023 Foundation Model Transparency Index indicates that developers of major open foundation models tend to be more transparent than their closed counterparts \citep{bommasani2023foundation}.

Still, model weights only make some types of transparency (\eg evaluations of risk) possible, but they do not guarantee such transparency will manifest.
More generally, model weights do not guarantee transparency on the upstream resources used to build the foundation model (\eg data sources, labor practices, energy expenditure) nor transparency on the downstream impact of the foundation model (\eg affected markets, adverse events, usage policy enforcement).
Such transparency can help address prominent societal concerns surrounding bias \citep{birhane_into_2023}, privacy \citep{ippolito_preventing_2023}, copyright \citep{henderson2023foundation, lee2023talkin, longpre2023data}, labor \citep{perrigo2023openai, hao2023cleaning}, usage practices \citep{narayanan2023transparencyreports}, and demonstrated harms \citep{guha2023ai}.

\textbf{Mitigating monoculture and market concentration.} 
\textit{Greater customizability mitigates the harms of monoculture and broader access reduces market concentration.}

Foundation models function as infrastructure for building downstream applications, spanning market sectors \citep{bommasani2021opportunities, bommasani2023ecosystem, vipra2023concentration, cma2023ai}.
By design, they contribute to the rise of algorithmic monoculture \citep{kleinberg2021monoculture, bommasani2022homogenization}: many downstream applications depend on the same foundation model.
Monocultures often yield poor societal resilience and are susceptible to widespread systemic risk: consider the Meltdown and Spectre attacks, which led to massive security risks because of the widespread dependence on Intel and ARM-based  microprocessors~\cite{kocher_spectre_2018,lipp_meltdown_2018,staff_meltdown_2018}. 
Further, foundation model monocultures have been conjectured to lead to correlated failures \citep{bommasani2022homogenization} and cultural homogenization \citep{lee2022language, padmakumar2023does}.
Since open foundation models are more easily customized, they may yield more diverse downstream model behavior, thereby reducing the severity of homogeneous outcomes.

Broad access to model weights and greater customizability further enable greater competition in downstream markets, helping to reduce market concentration at the foundation model level from vertical cascading.
In the foundation model market, there are barriers to entry for low-resource actors in developing foundation models given their significant capital costs \citep{vipra2023concentration, cma2023ai}.
For example, training the Llama 2 series of models required 3.3 million GPU hours on NVIDIA A100-80GB GPUs \citep{touvron2023llama2}: at February 2024 cloud computing rates of \$1.8/GPU hour~\cite{lambda_gpu_2024},
training this model would cost around \$6 million.
Further, while open foundation models may increase competition in some regions of the AI supply chain, they are unlikely to reduce market concentration in the highly concentrated upstream markets of computing and specialized hardware providers~\citep{widder2023open}.

\begin{table*}[ht!]
  \centering
  \begin{adjustbox}{width=\textwidth}
    \begin{tabular}{p{4.8cm}|p{3cm}|p{0.9cm}p{0.9cm}p{0.9cm}p{0.9cm}p{0.9cm}p{1.3cm}}
      \textbf{Misuse risk} & \textbf{Paper} & \rot{Threat identification} & \rot{Existing risk (absent open FMs)} & \rot{Existing defenses (absent open FMs)} & \rot{Evidence of marginal risk} & \rot{Ease of defense} & \rot{Uncertainty/assumptions}\\
      \midrule
      Spear-phishing scams & \citet{hazell_large_2023} & \bcircle & \wcircle & \ecircle & \ecircle & \wcircle & \ecircle\\
      \rowcolor[gray]{0.9} 
      Cybersecurity risk & \citet{seger2023open} & \wcircle & \ecircle & \wcircle & \ecircle & \wcircle & \ecircle\\
      Disinformation & \citet{musser_cost_2023} & \bcircle & \wcircle & \ecircle & \ecircle & \wcircle & \bcircle\\
      \rowcolor[gray]{0.9} 
      Biosecurity risk & \citet{gopal_will_2023} & \bcircle & \ecircle & \wcircle & \ecircle & \ecircle & \ecircle\\
      Voice-cloning scams & \hyperlink{ovadya_reducing_2019}{Ovadya et al.} \citeyearpar{ovadya_reducing_2019} & \bcircle & \wcircle & \wcircle & \wcircle & \wcircle & \bcircle\\
      \rowcolor[gray]{0.9} 
      Non-consensual intimate imagery & \citet{lakatos_revealing_2023} & \bcircle & \wcircle & \ecircle & \wcircle & \wcircle & \ecircle\\
      Child sexual abuse material & \citet{thiel_generative_2023} & \bcircle & \bcircle & \bcircle & \bcircle & \bcircle & \bcircle\\
      \bottomrule
    \end{tabular}
  \end{adjustbox}
\caption{Misuse analyses of open foundation models assessed under our risk framework (\autoref{subsec:framework}).
\bcircle{} indicates the step of our framework is clearly addressed;  \wcircle{} indicates partial completion; \ecircle{} indicates the step is absent in the misuse analysis. Incomplete assessments do not indicate that the analysis in prior studies is flawed, only that these studies, on their own, do not show an increased marginal societal risk stemming from open foundation models. 
We provide more details for our assessment of each row in~\autoref{app:risk_assessment}. 
}
  \label{table:risk_assessment}
\end{table*}

\hypertarget{risks}{\section{Risks of Open Foundation Models}}
\label{sec:risks}
Technologists and policymakers have worried that open foundation models present risks, in particular, due to the inability to monitor, moderate, or revoke access.
We survey the literature on misuse vectors specifically associated with open foundation models, identifying biosecurity, cybersecurity, voice cloning scams, spear phishing, disinformation, non-consensual intimate imagery, and child sexual abuse material \citep{seger2023open,thiel_generative_2023,maiberg_civitai_2023}.\footnote{Some have also discussed that (open) foundation models may contribute to existential risk via speculative AI takeover scenarios, which we do not consider here.}
To understand the nature of these risks, we present a framework that centers \textit{marginal} risk: what additional risk is society subject to because of open foundation models relative to pre-existing technologies, closed models, or other relevant reference points? 

\subsection{Risk Assessment Framework}\label{subsec:framework}
To assess the risk of open foundation models for a specific misuse vector, we present a six-point framework.
Underpinning this is an emphasis on communicating assumptions and uncertainty: misuse vectors often involve complex supply chains and the capabilities of foundation models are rapidly evolving, meaning the balance of power between attackers and defenders can be unstable~\cite{shevlane_offense-defense_2020}. 

The risk framework enables precision in discussing the misuse risk of open foundation models and is based on the threat modeling framework in computer security~\cite{drake_threat_2021,Shostack2014ThreatModeling, 10177704, Seaman2022CyberThreat, drake_threat_2021}.
For example, without clearly articulating the marginal risk of biosecurity concerns stemming from the use of open (natural) language models, researchers might come to completely different conclusions about whether they pose risks: open language models can generate accurate information about pandemic-causing pathogens \citep{gopal_will_2023}, yet such information is publicly available on the Internet, even without the use of open language models \citep{guha2023ai}.\footnote{In addition, two recent studies found that access to language models does not significantly increase access to information required to carry out biosecurity attacks compared to Internet access~\cite{mouton_operational_2024,patwardhan_building_2024}. More importantly, access to information might not be a major barrier for carrying out such attacks---stronger interventions might lie downstream~\cite{batalis_can_2023}.}

\textbf{1. Threat identification.}
All misuse analyses should systematically identify and characterize the potential threats being analyzed~\cite{Shostack2014ThreatModeling, 10177704, Seaman2022CyberThreat, drake_threat_2021}. 
In the context of open foundation models, this would involve naming the misuse vector, such as spear-phishing scams or influence operations, as well as detailing the manner in which the misuse would be executed. 
To present clear assumptions, this step should clarify the potential malicious actors and their resources: individual hackers are likely to employ different methods and wield different resources relative to state-sponsored entities. 

\textbf{2. Existing risk (absent open foundation models).}
Given a threat, misuse analyses should clarify the existing misuse risk in society. 
For example, \citet{seger2023open} outline the misuse potential for open foundation models via disinformation on social media, spear-phishing scams over email, and cyberattacks on critical infrastructure. 
Each of these misuse vectors already are subject to risk \textit{absent} open foundation models: understanding the pre-existing level of risk contextualizes and baselines any new risk introduced by open foundation models.

\textbf{3. Existing defenses (absent open foundation models).}
Assuming that risks exist for the misuse vector in question, misuse analyses should clarify how society (or specific entities or jurisdictions) defends against these risks.
Defenses can include technical interventions (\eg spam filters to detect and remove spear-phishing emails) and regulatory interventions (\eg laws punishing the distribution of child sexual abuse material). 
Understanding the current defensive landscape informs the efficacy, and sufficiency, with which new risks introduced by open foundation models will be addressed.

\begin{table*}[t!]
  \centering
  \begin{adjustbox}{width=\textwidth}
    \begin{tabular}{p{2.2cm}|p{9.3cm}|p{9.3cm}}
      \textbf{Framework step} 
      &
      \textbf{Cybersecurity}\newline Automated vulnerability detection 
      &
      \textbf{Non-consensual intimate imagery (NCII)} \newline Digitally altered NCII\\
      \midrule
      \parbox[t]{2cm}{Threat \\ identification}
      &
      Vulnerability detection tools can be used to automate the process of discovering software vulnerabilities. 
      Threat actors include individual hackers, small groups, or state-sponsored attackers. 
      
      &
      Digital tools can be used to alter images of people without their consent in sexually explicit ways. Threat actors are typically individuals or coordinated groups (such as on online platforms like Reddit or Telegram) creating imagery of people they know as well as public figures. \\
      \rowcolor[gray]{0.9} 
      Existing risk (absent open FMs)
      &
      Attackers benefit from the natural worst-case asymmetry in vulnerability detection: attackers need to exploit only a single effective vulnerability to succeed, whereas defenders must defend against all vulnerabilities to succeed.
      Existing risk is heavily influenced by the resources of the attacker: sophisticated attackers often make use of automated vulnerability detection tools in attack design.
      Fuzzing tools have long been used to find vulnerabilities in software~\cite{takanen_fuzzing_2008}, as have tools like Metasploit, a free penetration testing framework that can aid automated vulnerability detection~\cite{kennedy2011metasploit}. 
      MITRE's Adversarial Threat Landscape for Artificial-Intelligence Systems, a cybersecurity threat matrix for adversarial machine learning, includes many techniques that make use of closed foundation models and other types of machine learning models to detect vulnerabilities~\cite{mitre2021atlas}. 
      &
      Photoshop has long been used to create digitally altered NCII~\cite{broughton_tennessee_2009}. In the last decade, tools to create NCII using face swapping and other rudimentary ML techniques have become popular~\cite{widder_limits_2022}. A telegram bot that used such techniques was used to generate over 100,000 sexualized images of women~\cite{ajder_automating_2020}. Digitally altered NCII and also be used to extort victims~\cite{joshi_they_2021,satter_fbi_2023}, in addition to its emotional and psychological tolls~\cite{roberts_fake_2019,scott_deepfake_2020,hao_deepfake_2021}. \\
      \parbox[t]{2cm}{Existing \\ defenses (absent open FMs)}
      &
      Cybersecurity defenses often adopt defense-in-depth strategies, where defenses are layered to ensure an exploit based on an unaddressed vulnerability in one layer does not affect other layers of defenses~\cite{kuipers_control_2006}.
      Within the vulnerability detection setting, defenders can preemptively use vulnerability detection tools to detect and patch security threats, again dependent on their access to resources.
      Incentive strategies, such as bug bounties, can tilt the offense-defense balance in favor of defense to some extent by incentivizing bug finders (hackers, security researchers, firms) to report vulnerabilities. 
      &
      The software for creating digitally altered NCII can run on consumer-grade devices and has proliferated widely. There are efforts to reduce the use of such tools for creating NCII in open source communities~\cite{widder_limits_2022}, but these efforts are unlikely to be sufficient since there are several mechanisms for accessing the software. However, online platforms where NCII is distributed, such as social media platforms, can take steps to curb its spread~\cite{thiel_online_2020}. For example, a nonprofit called Stop NCII coordinates takedowns of known NCII across online platforms~\cite{mortimer_stopnciiorg_2021}.\\
      \rowcolor[gray]{0.9} 
      Evidence of marginal risk of open FMs 
      &
      We are unaware of existing evidence that malicious users have successfully used open foundation models to automate vulnerability detection. 
      Dark web advertisements for tools exist, claiming to facilitate automated vulnerability detection, but it is unclear if these products rely on open FMs~\cite{amos_what_2023}. 
      In considering marginal risks relative to closed foundations, while closed foundation models can be better monitored for misuse, it is not clear if such uses will be reliability identified.
      Namely, using a closed foundation model for vulnerability detection is not necessarily misuse, which introduces a nontrivial classification problem of distinguishing between legitimate and malicious uses of closed foundation models for automated vulnerability detection (see \cref{fig:chatgpt-vuln,fig:llama-vuln}).  
      &
      Over the last two years, open FMs have been used for creating vast amounts of digitally altered NCII. Compared to previous tools for creating sexualized imagery, open FMs can be fine tuned to create sexualized images of specific people~\cite{maiberg_inside_2023}. Compared to using tools like Photoshop, once such a fine-tuned model is made available, it is much easier for nonexperts to use these tools. While developers of closed FMs can enforce guardrails on the use of their text-to-image models for creating NCII, such guardrails on open FMs can be easily circumvented. There have been several real-world incidents involving the use of open FMs for creating NCII, leading to clear, demonstrated harm~\citep{llach2023naked, canas2023illegal, kaspersky2023how}.\\
      \parbox[t]{2cm}{Ease of \\ defense}
      &
      Similar to previous waves of automated vulnerability detection, LLMs can be incorporated into the information security toolkit to bolster defense. For example, \citet{liu_ai-powered_2023} show how LLMs can expand the coverage of a popular fuzzing tool called OSS-Fuzz. Foundation models can be used to monitor signals from deployed software systems for signs of active exploits as well. Google has made use of LLMs in its popular malware detection platform VirusTotal, using models to help explain the functionality of malware contained in a particular file~\cite{quintero2023vt}. Defense in depth will continue to remain important in aiding defense. Regardless of whether the model used for automated vulnerability detection is open or proprietary, signals and the ability to analyze them at machine scale and speed differentially supports defenders because of better access to the systems. 
      &
      Open FMs used to create NCII require few resources to run---indeed, many prominent text-to-image models can run on an iPhone or MacBook. As a result, non-proliferation of these models is generally not feasible. In contrast, crackdowns on the distribution of specifically tailored models for creating NCII is feasible and warranted, as is distribution of the content~\cite{gorwa_moderating_2023,maiberg_civitai_2023}. There are several legislative proposals to penalize the creation and distribution of digitally altered NCII, though given that channels for the spread of NCII can be anonymous or end-to-end encrypted, the efficacy of such legislation remains to be seen~\cite{illinois_general_assembly_full_2023,saliba_sharing_2023,reid_deepfake_2020,kocsis_deepfakes_2021,hao_deepfake_2021,siddique_sharing_2023}.
      \\
      \rowcolor[gray]{0.9} 
      \parbox[t]{2cm}{Uncertainty and \\ assumptions}
      &
      The analysis of marginal risk and ease of defense assumes that defenders will continue to have better access to state-of-the-art vulnerability detection tools, including those based on open FMs. It also assumes investment by defenders in using these tools to update their infosec practices and that the offense-defense balance will not change dramatically as the capabilities of models improve. 
      &
      Technical solutions for curtailing the use of already existing models to create NCII are hard or impossible. Even if future models can have robust technical safeguards, already-released models will continue to be misused. Even if downstream providers take steps to moderate digitally altered NCII, misuse can happen on smaller (anonymous/end-to-end encrypted) platforms or on the dark web by malicious users. \\
      \bottomrule
    \end{tabular}
  \end{adjustbox}
  \caption{Instantiation of our risk analysis framework for cybersecurity (automated vulnerability detection) and non-consensual intimate imagery (digitally altered NCII).}
  \label{table:framework_instantiation}
\end{table*}

\textbf{4. Evidence of marginal risk of open FMs.}
The threat identification, paired with an analysis of existing risks and defenses, provides the conceptual foundation for reasoning about the risks of open foundation models.
Namely, subject to the status quo, we can evaluate the \textit{marginal risk} of open foundation models.
Being aware of existing risk clarifies instances where open foundation models simply duplicate existing risk (\eg an open language model providing biological information available via Wikipedia).
Similarly, being aware of existing defenses clarifies instances where open foundation models introduce concerns that are well-addressed by existing measures~\citep[\eg, email and OS-based filters detecting spear-phishing emails, whether human or AI-generated;][]{craigmarcho_ie7_2007, apple_support_safely_2023, google_email_2023}.
Conversely, we can identify critical instances where new risks are introduced~\citep[\eg fine tuning models to create non-consensual intimate imagery of specific people; see \cref{table:framework_instantiation};][]{maiberg_inside_2023} or where existing defenses will be inadequate~\citep[\eg AI-generated child sexual abuse material may overwhelm existing law enforcement resources;][]{harwell_ai-generated_2023}.

Further, the marginal risk analysis need not only be conducted relative to the status quo, but potentially relative to other (possibly hypothetical) baselines.
For example, understanding the marginal risk of open release relative to a more restricted release (\eg API release of a closed foundation model) requires reasoning about the relevant existing defenses for said restricted release.
This perspective ensures greater care is taken to not assume that closed releases are intrinsically more safe and, instead, to interrogate the quality of existing defenses \citep[\eg the fallibility of existing API safeguards;][]{qi2023finetuning}.

\textbf{5. Ease of defending against new risks.}
While existing defenses provide a baseline for addressing new risks introduced by open foundation models, they do not fully clarify the marginal risk.
In particular, new defenses can be implemented or existing defenses can be modified to address the increase in overall risk.
Therefore, characterizations of the marginal risk should anticipate how defenses will evolve in reaction to risk: for example, (open) foundation models may also contribute to such defenses (\eg the creation of better disinformation detectors; \citet{zellers2019neuralfakenews} or code fuzzers; \citet{liu_ai-powered_2023}).

\textbf{6. Uncertainty and assumptions.}
Finally, it is imperative to articulate the uncertainties and assumptions that underpin the risk assessment framework for any given misuse risk. 
This may encompass assumptions related to the trajectory of technological development, the agility of threat actors in adapting to new technologies, and the potential effectiveness of novel defense strategies. 
For example, forecasts of how model capabilities will improve or how the costs of model inference will decrease would influence assessments of misuse efficacy and scalability.

Using our risk assessment framework, we assess past studies that span different risk vectors in \autoref{table:risk_assessment}.
We find that the risk analysis is incomplete for six of the seven studies we analyze.
To be clear, incomplete assessments do not necessarily indicate that the analysis in prior studies is flawed, only that these studies, on their own, are insufficient evidence to demonstrate increased marginal societal risk from open foundation models. 

In \autoref{table:framework_instantiation}, we instantiate the framework for two misuse risks, providing preliminary analyses of cybersecurity risks stemming from automated vulnerability detection and the risk of digitally altered NCII. 
For the former, we find that the current marginal risk of open foundation models is low and that there are several approaches to defending against the marginal risk, including using AI for defense. 
For the latter, open foundation models pose considerable marginal risk at present, and plausible defenses seem hard. 
Note that these are not the only risks from foundation models~\cite{barrett_identifying_2023}---for example, the creation of malware is another cybersecurity risk that requires separate analysis---yet when researchers talk about cybersecurity risks of open foundation models, they often club together different threats. 
This illustrates how the framework helps clarify the points of contention in debates on open foundation models. Critically, while many of the same properties of open foundation models are relevant for analyzing different misuse vectors (such as the inability to revoke access), the risk assessment framework helps introduce specifics that differentiate the misuse vector, for instance, by pointing out elements of the misuse supply chain where risk is better addressed. 

As the capabilities of foundation models (including open models) improve, the risk assessment framework can guide analyses of societal risks from increasing capability by providing a grounded analysis of whether model releases bring about increased marginal risk to society. Still, it is important to note the limitations on the scope of the framework's applicability. First, while the risk assessment framework can help clarify the societal risks of releasing a foundation model openly, note that it is not a complete framework for making release decisions since it does not provide a mechanism for trading the marginal benefits of openly releasing models against the marginal risk, nor does it look at the opportunity cost of \textit{not} releasing a model openly. Second, while the framework allows an evaluation of the risk of releasing models openly for known risks (such as cybersecurity, biosecurity etc.), it does not account for \textit{unknown unknowns}---risks that we have no prior understanding of. 
Third, there could be a number of coordination issues among actors for figuring out when to release models---for example, to reduce the risk of NCII, open model developers would need to coordinate with social media platforms as well as other downstream platforms like CivitAI (see \cref{table:framework_instantiation}). While the framework allows us to identify such opportunities, it does not automatically bring about the coordination of these actors.
Overall, while the framework improves the precision, rigor, and completeness of risk assessment, we expect other approaches to analyzing risk will be needed for addressing these limitations.

\hypertarget{recommendations}{\section{Recommendations and calls to action}}
\label{sec:discussion}

Armed with a clearer conceptualization of the benefits and a framework for assessing the risks of open foundation models, we make the following recommendations to (i) AI developers, (ii) researchers investigating AI risks, (iii) policymakers, and (iv) competition regulators.

\textbf{AI developers.}
In contrast to closed foundation models, which are usually treated by their developers and their users with product safety expectations, open foundation models have less clear safety expectations.
In particular, the division of responsibility for safety between the developer and user of an open foundation model is unclear and lacks established norms.
Consequently, developers of open foundation models should be transparent about both the responsible AI practices they implement and the responsible AI practices they recommend or delegate to downstream developers or deployers.
In turn, when downstream developers are procuring foundation models, they should consider which responsible AI measures have already been implemented (and their efficacy if measured) and, accordingly, implement or bargain for responsible AI practices.
This would help ensure that responsible AI practices do not fall through the cracks as providers of downstream AI applications leverage open foundation models from other upstream providers.

\textbf{Researchers investigating AI risks.}
Our preliminary analysis of the misuse risk of open foundation models reveals significant uncertainty for several misuse vectors due to incomplete or unsatisfactory evidence.
In turn, researchers investigating AI risks should conduct new research to clarify the marginal risks for misuse of open foundation models.
In particular, in light of our observations regarding past work, greater attention should be placed on articulating the status quo, constructing realistic threat models (or arguments for why speculative threat models yield generalizable evidence), and considering the full supply chain for misuse. 
\pl{I would say something about scaling laws for risks - or minimally, taking into account that the technology is rapidly progressing and offering risks assessments that anticipate future model capabilities.}

\textbf{Policymakers.}
Government funding agencies should ensure that research investigating the risks of open foundation models is sufficiently funded while remaining appropriately independent from the interests of foundation model developers \citep{lucas_letter_2023}.
Once the uncertainty around specific misuse vectors is reduced (including via improved tracing of downstream model usage), and if the marginal risks are shown to be significant enough to warrant concern, further policy interventions (\eg hardening downstream attack surfaces) can be considered.
Policymakers should also proactively assess the impacts of proposed regulation on developers of open foundation models.
In particular, some policy proposals impose high compliance burdens for these developers, and such policies should only be pursued with sufficient justification of the adverse effect on the open foundation model ecosystem.
Policies that place obligations on foundation model developers to be responsible for downstream use are intrinsically challenging, if not impossible, for open developers to meet.
If recent proposals for liability \citep{blumenthal_bipartisan_2023} and watermarking \citep{EO14110,chinese_national_information_security_standardization_technical_committee_basic_2023, g7_hiroshima_summit_hiroshima_2023} are interpreted strictly to apply to foundation model developers, independent of how the model is adapted or used downstream, they would be difficult for open developers to comply with \cite{bommasani_considerations_2023}, since these developers have little ability to monitor, moderate, or prohibit downstream usage.

\textbf{Competition regulators.}
Significant theoretical benefits of open foundation models relate to their potential to catalyze innovation, distribute power, and foster competition. 
With this in mind, the magnitude of these economic benefits is largely undocumented in the absence of large-scale economic analyses or market surveillance.
For example, many benefits hinge on open foundation models meaningfully expanding consumer choice and reducing costs. 
If factors such as differences in model quality are the more direct causes of the adoption of specific foundation models, these purported benefits may not manifest.
In turn, competition regulators should invest in measuring the benefits of foundation models and the impact of openness on those benefits.
In particular, the UK's Competition and Markets Authority has begun such work \citep{cma2023ai}, which would be bolstered by parallel efforts across other jurisdictions. 
\hypertarget{conclusion}{\section{Conclusion}}
\label{sec:conclusion}

Open foundation models are controversial due to fundamental philosophical disagreements, fragmented conceptual understanding, and poor empirical evidence.
Our work aims to rectify the conceptual confusion by clearly defining open foundation models, identifying their distinctive properties, and clarifying their benefits and risks.
While it is unlikely that certain underlying philosophical tensions will ever be resolved, especially when inextricably intertwined with the incentives of different actors in the AI space, we encourage future work to address today's deficits in empirical evidence.
Overall, we are optimistic that open foundation models can contribute to a vibrant AI ecosystem, but realizing this vision will require significant action from many stakeholders.
 
\section*{Acknowledgements}
We thank Steven Cao, Nicholas Carlini, Zico Kolter, Ellie Evans, Helen Toner, and Ion Stoica for extensive feedback on this work. 
We are grateful to the participants of the Stanford Workshop on the Governance of Open Foundation Models and the Princeton-Stanford Workshop on Responsible and Open Foundation Models for their feedback and input.
This work was supported in part by the AI2050 program at Schmidt Futures (Grant G-22-63429), the Patrick J. McGovern Foundation and the Hoffman-Yee grant program of the Stanford Institute for Human-Centered Artificial Intelligence.

\bibliography{references, references-1, references-2, all}
\bibliographystyle{icml2024}

\newpage
\appendix
\renewcommand{\thefigure}{A\arabic{figure}}
\renewcommand{\thetable}{A\arabic{table}}

\setcounter{figure}{0}
\setcounter{table}{0}
\onecolumn
\section*{Appendix}

\begin{table*}[t!]
\centering
\footnotesize
\renewcommand{\arraystretch}{1.5}
\begin{adjustbox}{width=\textwidth}
\begin{tabular}{p{3.5cm} p{3cm} p{10cm}}
\toprule
\textbf{Paper} & \textbf{Domain} & \textbf{Summary of Research} \\
\midrule

\citet{yang2023shadow} & Safety and security & Safety alignment can be subverted with minimal finetuning.
\\ \rowcolor[gray]{0.9}
\citet{choi2023tools} & Privacy and security & Assessing the ability to verify a model's training data.
\\
\citet{patil2023can} & Privacy and security & Methods to prevent sensitive information extraction attacks.
\\ \rowcolor[gray]{0.9}
\citet{zou2023universal} & Safety and alignment & Adversarial attacks that transfer from open models to black-box, closed models.
\\ 
\citet{kirchenbauer_watermark_2023} & Content provenance & Watermarking methods for LLMs
\\ \rowcolor[gray]{0.9}
\citet{dettmers2023qlora} & Training efficiency & Efficient training with quantized low rank adapters.
\\
\citet{longpre2023pretrainer} & Toxicity and bias & Understanding the effects of ``quality'' filters on model toxicity and performance.
\\ \rowcolor[gray]{0.9} 
\citet{li2023structural} & Brain imaging analysis & Comparing the representations of sequences in LLMs and neural response measurements.
\\ 
\citet{han2023medalpaca} & Medical applications & Training medical application models.
\\ \rowcolor[gray]{0.9}
\citet{wang2022interpretability} & Interpretability & Explaining model behaviors in terms of their internal components.
\\ 
\citet{tian2023joma} & Architecture analysis & Understanding training dynamics in multi-layer Transformer architectures. \\
\bottomrule
\end{tabular}
\end{adjustbox}
\caption{A non-comprehensive list of research that uses open foundation models, organized by the research domain. The example works are enabled with GPT-2 \citep{Radford2019LanguageMA}, LLaMA \citep{touvron2023llama}, Llama-2 \citep{touvron2023llama2}, Pythia \citep{biderman2023pythia}, GPT-J \citep{wang2021gpt}, GPT-NeoX \citep{black2022gpt}, Bloom \citep{scao2022bloom}, and OPT \citep{zhang2022opt}. Based on data compiled by \citet{biderman_good_2023}.
}
\label{tab:science}
\end{table*}

\begin{figure}[h!]
\centering
\begin{subfigure}[t]{0.49\textwidth}
  \centering
  \begin{tikzpicture}
    \node[inner sep=0] (image) at (0,0) {\includegraphics[height=11.5cm]{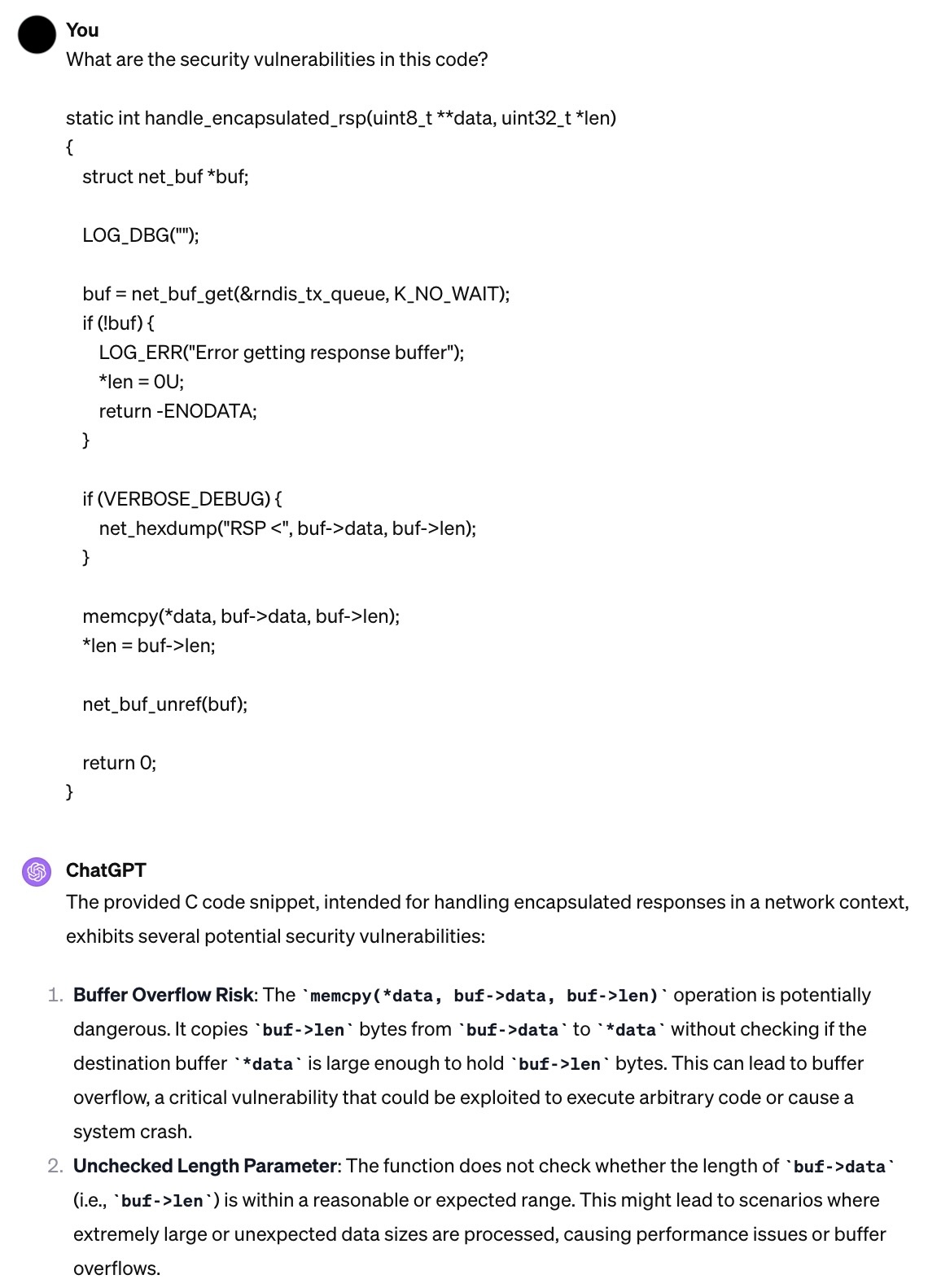}};
    \draw[black, rounded corners, thin] (image.south west) rectangle (image.north east);
  \end{tikzpicture}
  \caption{When prompted with the code from a recent security vulnerability~\cite{flavio_ceolin_two_2023}, ChatGPT is able to diagnose the issue.}
  \label{fig:chatgpt-vuln}
\end{subfigure}
\hfill
\begin{subfigure}[t]{0.49\textwidth}
  \centering
  \begin{tikzpicture}
    \node[inner sep=0] (image) at (0,0) {\includegraphics[height=11.5cm]{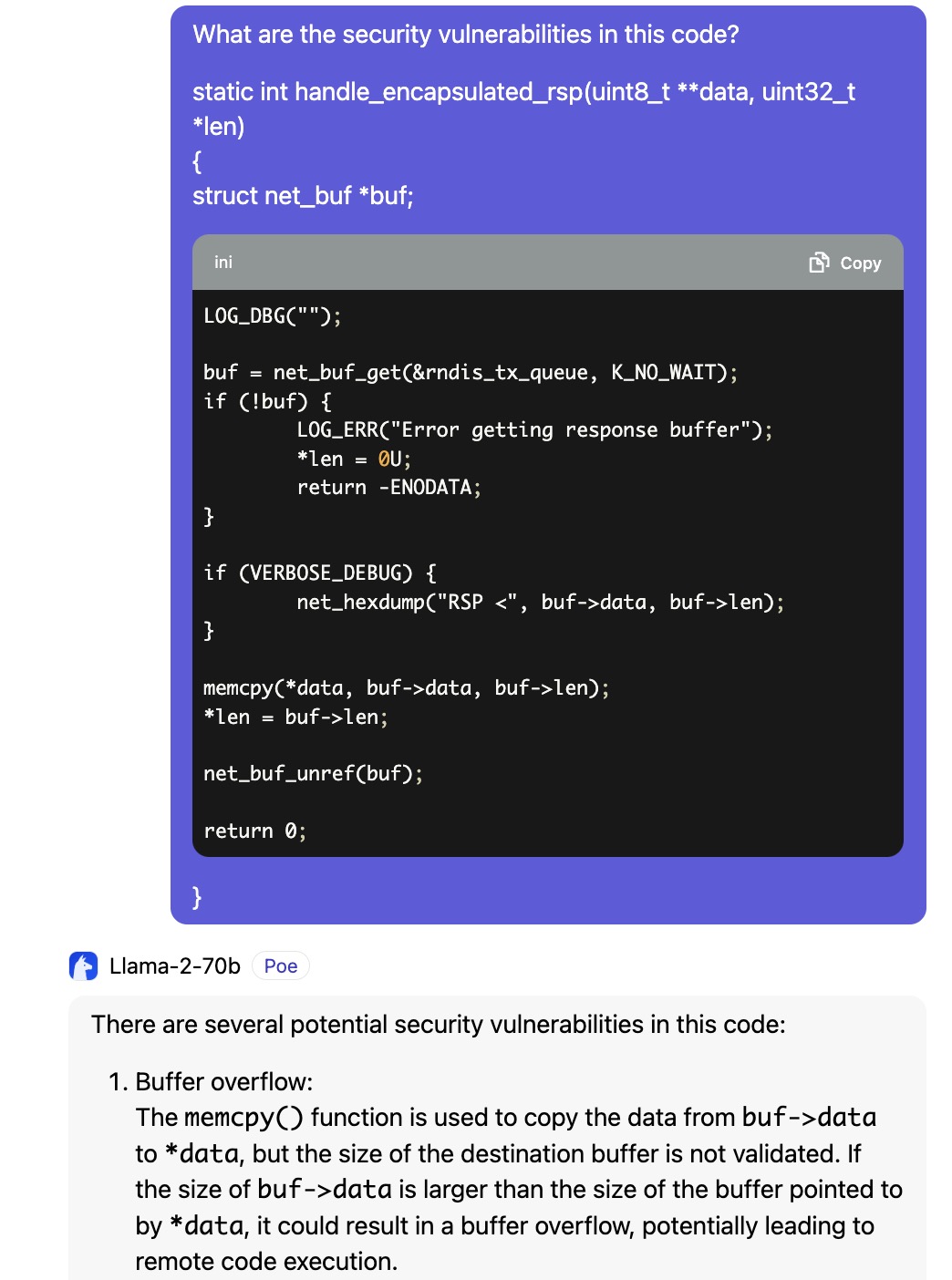}};
    \draw[black, rounded corners, thin] (image.south west) rectangle (image.north east);
  \end{tikzpicture}
  \caption{Llama-2 is similarly able to spot the buffer overflow issue.}
  \label{fig:llama-vuln}
\end{subfigure}
\caption{Comparison of ChatGPT and Llama-2 in identifying vulnerabilities}
\end{figure}

\section{A brief history of open foundation models}
\label{sec:prior work}

In the 2010s, the advent of sophisticated generative image models introduced the concern of deepfakes that could misinform viewers \citep{paris_deepfakes_2019, chesney_deep_2018, widder_limits_2022, cole_ai-assisted_2017}.
These models facilitated significant impersonation, political misinformation, and non-consensual intimate imagery (NCII). 
For example, a Telegram bot was used to generate over a hundred thousand nude images of women~\cite{solsman_deepfake_2020}. 
These models were not foundation models; they relied on more rudimentary algorithms, such as swapping faces in images.

In the late 2010s, foundation models gave rise to powerful generative capabilities for language \citep{radford2018improving}, and new misuse concerns came to the fore.
In February 2019, OpenAI announced the GPT-2 series of models \citep{radford_language_2019,openai_better_2019}: four language models ranging between 124 million and 1.5 billion parameters. 
Primed by several concerns --- especially the potential for large-scale disinformation \citep{solaiman_release_2019} --- OpenAI opted for a staged release where models of increasing size were openly released from February to November 2019. 
While the company did not ultimately find evidence of model misuse during the staged release \citep{openai_gpt-2_2019}, this process brought attention to the nexus between release and misuse \citep{bommasani2021opportunities, sastry2021release, shevlane2022structured, liang2022norms, solaiman2023gradient}.

Since the release of GPT-2, hundreds of foundation models have been released by various actors adopting different release strategies \citep{bommasani2023ecosystem}.
The open release of Stable Diffusion in August 2022 was particularly salient, as it was one of the first text-to-image models to be available widely outside the research community.
However, because the model weights were shared publicly, users easily circumvented filters that Stability AI implemented to prevent the generation of not safe for work (NSFW) imagery. 
As a result, AI-generated pornography based on Stable Diffusion offshoots quickly spread across the internet, including images resembling real people generated without their consent \citep{maiberg_inside_2023}.

Meta's release of its LLaMA language models \citep{touvron2023llama} in March 2023 marked another significant event in the trajectory of open foundation models.
LLaMA was released via a form that allowed researchers to download model weights by accepting a license for non-commercial use. 
But the model weights were quickly leaked, leading to concerns that the highly capable model could facilitate misuse. 
U.S. Senators Hawley and Blumenthal sent a letter to Meta CEO Mark Zuckerberg expressing concerns about Meta’s release strategy~\citep{blumenthal2023meta}.

\section{Risk assessment framework} 
\label{app:risk_assessment}

We surveyed the literature on misuse vectors specifically associated with open foundation models, identifying biosecurity, cybersecurity, voice cloning scams, spear phishing, disinformation, non-consensual intimate imagery, and child sexual abuse material. For each of these misuse vectors, we chose a past study that analyzed the misuse vector and discussed or analyzed openness. Note that these studies did not need to compare open models to closed ones or other existing technologies, such as the internet, to be considered. And the lack of evidence in one of the risk assessment framework elements does not mean that the study's analysis is flawed; it only means that the study does not present adequate evidence of an increased marginal societal risk of open foundation models. Below, we present our justifications for the scores for each study in \Cref{table:risk_assessment}. 

\subsection{Spear-phishing scams~\cite{hazell_large_2023}}

\textbf{Threat identification \bcircle}
The threat is specified clearly---scams due to emails that appear personalized. The threat actor is specified as low-skill actors who can focus on high-level planning and outsource the email writing to LLMs. (See sections 4 and 5.)

\textbf{Existing risk (absent open FMs) \wcircle}  
There is some characterization of existing risk (e.g., examples of successful phishing scams), but no broader analysis of how societally impactful spear-phishing scams currently are. This is essential to understanding the arguments' scope of overall and marginal risk. (See sections 2 and 4.)

\textbf{Existing defenses (absent open FMs) \ecircle}  
While there is a brief discussion of some defenses, there is little analysis of how effective protections such as those built into modern operating systems or email services are at preventing such spear-phishing scams. (See, for example, section 7.2.)

\textbf{Evidence of marginal risk \ecircle}
There is no analysis of how open FMs can be used to circumvent existing defenses built into operating systems or email services. (See, for example, section 5.3.).

\textbf{Ease of defense \wcircle}
The paper discusses AI-based defenses (e.g., using Sec-PaLM), and discusses how foundation models should be governed given the findings of the study, which may provide a further defense. However, there is no analysis on how robust existing defenses are, and whether further defenses are needed. (See section 7.)

\textbf{Uncertainty/assumptions \ecircle}
The paper does not explicitly analyze assumptions underlying the analysis (e.g., existing defenses will fail; content creation is already cheap without AI models) or how these assumptions could fail in the real world. (See, for example, section 5.)

\subsection{Cybersecurity risk~\cite{seger2023open}}

\citet{seger2023open} outline several risks of releasing foundation models openly (including many of the other risks we look at---disinformation, scams etc.); here, we focus on their analysis of cybersecurity risks.

\textbf{Threat identification \wcircle}
The main threat model described is the creation of malware, in very broad terms. No clear threat actors are identified. (See section 3.1.1.)

\textbf{Existing risk (absent open FMs) \ecircle}  
The paper does not discuss how prevalent malware-based cyberattacks currently are, or their societal impact, or provide examples of the type of cyberattacks that are in scope. (See, for example, section 3.1.1.)

\textbf{Existing defenses (absent open FMs) \wcircle}  
Some examples of defenses (e.g., bug bounties) are provided, but there is no in-depth analysis of how helpful these defenses are against existing risks, excluding AI systems. (See section 4.1.3.)

\textbf{Evidence of marginal risk \ecircle}
The paper does not provide any evidence that open foundation models have contributed to cybersecurity incidents by facilitating the creation of malware. It does not compare against other baselines, such as finding similar information on the Internet or using closed models (e.g., via jailbreaks or fine tuning). (See, for example, sections 3.1 and 3.2.)

\textbf{Ease of defense \wcircle}
The paper briefly discusses bug bounties as a mechanism for hiring experts to bolster defense, but does not analyse how much interventions reduce marginal risk. It also acknowledges the arguments of open models advancing safety research, but does not assess alternative ways to mitigate marginal risk (aside from staged release) such as using LLMs as tools for cyberdefense. (See section 4.1.3.)

\textbf{Uncertainty/assumptions \ecircle}
The paper acknowledges that the offense-defense balance is tentative, but does not analyze it in the context of cybersecurity or articulate other core assumptions that affect the assessment of open and closed models, such as the resources available to the most concerning threat actors. (See, for example, sections 3.1.1 and 3.1.3.)

\subsection{Disinformation~\cite{musser_cost_2023}}

\textbf{Threat identification \bcircle}
The threat is identified clearly: propagandists can cheaply create content to conduct influence operations. (See sections 1 and 2.)

\textbf{Existing risk (absent open FMs) \wcircle}  
The paper gives some examples of preexisting risks of mass content generation (e.g., by the Chinese government and Russian agencies). It does not, however, analyze the societal impact of such influence operations beyond the number of content posts generated, or the effectiveness of current methods at influencing people at scale. (See section 2.)

\textbf{Existing defenses (absent open FMs) \ecircle}  
The paper gives makes some reference to efforts by social media companies and language model developers to defend against influence operations, but it does not analyze the effectiveness of these or other existing defenses. (See sections 3 and 4.) 

\textbf{Evidence of marginal risk \ecircle}
The paper gives evidence of cost reduction for the generation of content, but does not discuss existing assessments of the demand for such content, whether it is actually effective in the real world, or if open foundation models can be used to circumvent existing defenses. It also does not analyze the costs associated with distributing said content (e.g., maintaining user profiles, avoiding behavioral social media moderation etc.), which is likely to overshadow the cost of content creation in disinformation operations. (See, for example, sections 3 and 4.)

\textbf{Ease of defense \wcircle}
The paper analyzes the cost of avoiding monitoring on closed models compared to open ones. It does not analyze steps social media platforms could take such as behavioral or network-based moderation, improving captcha to detect bots, or other existing mechanisms already in use by social media platforms. (See section 5.)

\textbf{Uncertainty/assumptions \bcircle}
The paper articulates several assumptions built into the analysis, such as the type of actors using the models, the relative costs of open vs. closed models, and the cost of evading monitoring for closed foundation models. (See sections 4, 5 and 8.)

\subsection{Biosecurity risk~\cite{gopal_will_2023}}

\textbf{Threat identification \bcircle}
The threat is identified clearly: individuals without training could obtain and create pandemic-causing pathogens. (See page 2.)

\textbf{Existing risk (absent open FMs) \ecircle}  
The paper provides no analysis of how a rogue individual or non-state actor could engage in bioterrorism absent foundation models (or open foundation models). (See, for example, page 2.)%

\textbf{Existing defenses (absent open FMs) \wcircle}  
There is a brief analysis of limitations to rogue individuals' and non-state actors' ability to leverage information about materializing biorisks. The paper mentions deterring effects such as the lack of information about viruses, the number of researchers capable of assembling an influenza virus, and existing immunity from historical pathogens. But there is no mention of other existing notable defenses such as controls on procurement of raw materials or benchtop DNA synthesizers. Nor is there any discussion of the efficacy of these defenses. (See page 3.)

\textbf{Evidence of marginal risk \ecircle}
The paper gives no comparison to similar risks based on widely available information on the Internet. The discussion of closed models' safeguards is not substantiated by evidence; for example, the paper does not analyze the risks from closed models related to jailbreaking or fine tuning. (See, for example, page 6.)

\textbf{Ease of defense \ecircle}
The paper proposes defenses such as legal liability and market-based insurance. There is no evidence regarding the feasibility or effectiveness of these proposed defenses, especially against well resourced rogue actors. Similarly, there is no analysis of how effective existing defenses (such as controls on procurement) would be. (See, for example, pages 7 and 8.) %

\textbf{Uncertainty/assumptions \ecircle}
The paper does not analyze assumptions built into the analysis, such as how future capabilities would evolve; whether current models are already capable enough to be fine tuned on biological information and generate information on how to cause \textit{novel} pandemics (which increases marginal risk); or how information on the Internet could similarly aid attackers (which would lower marginal risk). (See, for example, page 7.)

\subsection{Voice cloning~\cite{ovadya_reducing_2019}}
\label{app:voice cloning}

\textbf{Threat identification \bcircle} The threat is identified clearly: financial scams due to voice cloning technology using machine learning. The paper identifies several potential threat actors, such as malicious users with or without ML expertise. (See section 2.2.)

\textbf{Existing risk (absent open FMs) \wcircle}  
There is some characterization of existing risk (such as financial crimes), but there is little broader analysis of the societal impact of voice cloning scams at present. (See sections 1 and 3.5.)

\textbf{Existing defenses (absent open FMs) \wcircle}  
There is some discussion of existing defenses implemented by companies (Synthesia, Lyrebird), but little analysis of their efficacy. There is no analysis of the efficacy of these defenses against preexisting risk. (See section 3.5.)

\textbf{Evidence of marginal risk \wcircle}
There is some discussion of the marginal risk of openness (e.g., related to reproducibility, modifiability, access ratchets), but minimal direct comparison to risks associated with closed models or other technologies that could similarly enable misuse. (See sections 2.2 and 4)

\textbf{Ease of defense \wcircle}
The paper analyzes several mechanisms for mitigating marginal risk (e.g. timing of release, what assets are released) but does not give a thorough account of how much these mitigations could address marginal risk. (See sections 3.3 and 3.4.)

\textbf{Uncertainty/assumptions \bcircle}
The paper articulates several assumptions built into its analysis such as the current state of open and closed models and how they are released. (See sections 2.2, 3.5, and 4.2-4.4.)

\subsection{Non-consensual intimate imagery (NCII)~\cite{lakatos_revealing_2023}}

\textbf{Threat identification \bcircle}
The threat is identified clearly: digitally altered NCII. Threat actors are people who do not necessarily require machine learning skills and can rely on easy-to-use interfaces in order to create NCII. (See page 1.)

\textbf{Existing risk (absent open FMs) \wcircle}  
The paper looks at comments referring users to NCII distribution sites before (and after) the widespread availability of open text-to-image foundation models and demonstrates that there are tens of millions of unique visitors to such sites, though it does not describe the risk of preexisting tools such as those for face swapping. (See pages 1 and 3.) 

\textbf{Existing defenses (absent open FMs) \ecircle}  
The paper does not examine existing technical or legal defenses against NCII that could be used to deter malicious users or prevent the spread of NCII. (See, for example, page 1.)

\textbf{Evidence of marginal risk \wcircle}
The paper outlines the risk stemming from open models, such as by showing that the increase in use of services for creating NCII is driven by open FMs, but does not compare the risk of open models to that of closed models or other digital technologies such as Photoshop. (See page 1.)

\textbf{Ease of defense \wcircle}
There is some discussion of downstream entities such as Paypal not providing services to platforms that enable the distribution of NCII, but there is no significant analysis of defenses against marginal risk posed by open foundation models specifically. (See page 5.) %

\textbf{Uncertainty/assumptions \ecircle}
There is no explicit discussion of the uncertainty or assumptions built into the analysis, such as the potential adaptation of defenses. (See, for example, page 1.)

\subsection{Child sexual abuse material (CSAM)~\cite{thiel_generative_2023}}

\textbf{Threat identification \bcircle}
The threat is clearly identified: the distribution of computer-generated CSAM (or CG-CSAM). It identifies threat actors as hobbyist groups and forums for sharing CSAM. (See section 1.)

\textbf{Existing risk (absent open FMs) \bcircle}  
The authors provide evidence that the existing prevalence of CG-CSAM is low. Based on an analysis of online forums dedicated to sharing CSAM, they find that less than 1\% of CSAM on online forums is photorealistic CG-CSAM. (See section 1.) %

\textbf{Existing defenses (absent open FMs) \bcircle}  
The paper outlines legal defenses as well as moderation by platforms (such as creating hashsets of known CSAM) as defenses against CSAM absent open FMs. (See sections 4 and 5.)

\textbf{Evidence of marginal risk \bcircle} 
The authors list several concerns from the use of open foundation models: re-victimization of abused children if models are trained on their likeness to create new images; liar's dividend---when perpetrators who possess real CSAM claim they possess CG-CSAM; increasing toll on moderators who need to filter through increasingly lifelike CG-CSAM to find examples of real abused children; and the increased burden on enforcement agencies from a surge of CG-CSAM. The paper also provides evidence that most of CG-CSAM is based on the Stable Diffusion series of models, which are openly available; points out that watermarking or monitoring the use of Stable Diffusion is hard because of open weights; and discusses how such watermarking has been disabled. (See sections 4 and 5.)

\textbf{Ease of defense \bcircle}
The paper discusses several defenses (and analyses their shortcomings), including: developers removing known CSAM from training data; methods to create persistent watermarks that increase barriers to entry (such as Stable signature); legal defenses against photo-realistic CSAM; and low-tech defenses such as using EXIF data to differentiate CG-CSAM. (See sections 4, 5 and 6.)

\textbf{Uncertainty/assumptions \bcircle}
The paper points out the scope for better models in increasing risk; 
the assumption about the difficulty of watermarking open models;
and the uncertainty about the legal status of CG-CSAM. (See sections 3, 4 and 5.)

\end{document}